\documentclass[a4paper, 12pt, twoside]{article} 
\usepackage{lmodern}

\usepackage[T1]{fontenc}
\usepackage{inputenc}
\usepackage[a4paper]{geometry}
\usepackage{color}
\usepackage{babel}
\usepackage{amsmath}
\usepackage{amsthm}
\usepackage{amssymb}
\usepackage{mathtools}
\usepackage{dsfont}
\usepackage{stmaryrd}
\usepackage{graphicx}
\usepackage{setspace}
\usepackage{esint}
\usepackage{caption}
\usepackage{subcaption}
\usepackage{rotating}
\usepackage{booktabs}
\usepackage{tikz}
\usetikzlibrary{fit,positioning,calc}
\tikzset{
  font={\fontsize{11pt}{12}\selectfont}}

\usepackage[unicode=true,bookmarks=true,bookmarksnumbered=false,bookmarksopen=false,breaklinks=false,pdfborder={0 0 0},pdfborderstyle={},backref=page,colorlinks=true]{hyperref}
\hypersetup{linkcolor=blue,citecolor=blue, urlcolor = blue}

\usepackage{babel}
\usepackage{fullpage}

\usepackage{xcolor}
\usepackage{mhequ}

\usepackage[normalem]{ulem}

\usepackage[super,sort&compress]{natbib}
\setcitestyle{comma,numbers,super,open={},close={}}

\newcommand{\be}{\begin{equs}}
\newcommand{\ee}{\end{equs}}
\newcommand{\covid}{\textsc{covid-19}}
\newcommand{\bx}{\mathbf{x}}
\newcommand{\bbeta}{\boldsymbol{\beta}}

\title{Dynamic modeling of the Italians' attitude towards Covid-19}
\date{}

\author{Emanuele Aliverti\thanks{Department of Economics, University  Ca' Foscari Venezia, emanuele.aliverti@unive.it},
Massimilano Russo\thanks{
Division of Pharmacoepidemiology and Pharmacoeconomics, Department of Medicine, Brigham and Women’s Hospital and Harvard Medical School,  mrusso@bwh.harvard.edu}}

\begin{document}

\maketitle

\begin{abstract}
\setstretch{1.0}
We analyze repeated cross-sectional survey data collected by  the Institute of Global Health Innovation,  to characterize the perception and behavior of the Italian population during the \textsc{Covid-19} pandemic, focusing on the period that spans from April 2020 to July 2021. 
To accomplish this goal, we propose a Bayesian dynamic latent-class regression model, that accounts for the effect of sampling bias including survey weights into the likelihood function.
According to the proposed approach, attitudes towards \textsc{covid-19} are described via ideal behaviors that are fixed over time,  corresponding to different degrees of compliance with spread-preventive measures.
The overall tendency toward a specific profile dynamically changes across survey waves via a latent Gaussian process regression, that adjusts for subject-specific covariates.
We illustrate the evolution of Italians' behaviors during the pandemic, providing insights on how the proportion of ideal behaviors has varied  during the phases of the lockdown, while  measuring the effect of age, sex, region and employment of the respondents on the attitude toward \covid{}.
\end{abstract}

\setstretch{1.0}

\setstretch{1.5}
\section{Introduction}

The outbreak of the \covid{} pandemic has impacted our world, with more than $250$ million infected people, and more than $5$ million deaths at December 2021.~\citep{who:2020}
From an economic perspective, the \covid{} outbreak and the national measures to contain the spread of  the disease lead to severe economic recessions in many sectors, such as tourism, accommodation, and food services.~\citep{euReport:2020}  
Psychological and social effects on the population are less immediate to quantify, but some preliminary results  suggest that the pandemic has affected also these aspects of human life.~\citep{cullen:2020,balkhi:2020, apa:2020}

In this article, we study the evolution of behaviors in compliance with some measures that prevent the spread of \covid{} (see Table~\ref{tab:dataset}). We focus on the Italian population, that represents an interesting case study, because Italy was the first European country to introduce a national lockdown to limit the spread of \covid{}, imposing behavioral changes in the population. 
The adopted measures are likely to have had a psychological and social effect proportionate with their severity and duration. In fact, some early reports, based on online surveys, suggest an increased level of distress, anxiety, and fear.~\citep{forte:2020,moccia:2020,motta:2020} 
Monitoring these aspects, quantifying their evolution over time, and characterizing their impact on individuals is of primary importance to evaluate the wellbeing of a population. 
Similarly, it is of interest to evaluate the compliance with \covid{} policies, that can largely depend on personal status.~\citep{carlucci:2020}

Variations of personal routines and  practices have been documented; for example, the reduction of social activities and gatherings and the changes in mobility patterns, such as the increase of work-from-home practices and the reduction of public transport use.~\cite{google,bavadekar:2020}
These behaviors reflect the compliance with the government regulations, as well as the internalization of different recommendations to reduce the spread of the virus, that changed day-to-day life.
However, compliance with spread reducing measures depends upon personal conditions, psychological status, and many other personal factors,~\citep{wolf:2020} and it is  likely to change during different stages of the pandemic. 

During the first phase of the pandemic (up to June 2020), there is evidence that the Italian population has scrupulously followed government measures~\citep{graffigna:2020,barari:2020}; however, to the best of our knowledge, no analysis has been performed to assess if such compliance is constant over time, across socio-demographic groups, or proportional to the severity of the measures adopted by the Italian government throughout the pandemic.  To quantify these aspects,  we analyze survey data provided by the Institute of Global Health Innovation (\textsc{ighi}) at the Imperial College of London, in collaboration with the company YouGov, \citep{jones2020imperial} described in Section~\ref{sec:data}.

We describe the attitude of the Italian population toward \covid{} throughout the pandemic with a dynamic Bayesian latent class regression model.~\citep{lazarsfeld:1950, dayton:1988}
Such models assume that the population is composed of $H$ groups corresponding to ideal behaviors, that can represent different attitudes towards \covid{}. 
At any given time, each subject composing the population is associated with one of these ideal profiles, and the categorical variables characterizing behaviors are modeled as conditionally independent given the profile memberships.
Latent class models are conceptually simple and have been used as a building block for several methods to analyze categorical variables; for example, in problems including survey weights, \citep{vermunt:2007,patterson:2002} when interest is on characterizing temporal dependence across contingency tables \citep{kunihamaAndDunson:2013} or differences among groups of subjects.~\citep{russoEtAl:2018} See also Chapters 9 and 11.5 of
\textit{Handbook of mixture analysis}\cite{mixture:2019} for further references.

The paper is structured as follows:  Section~\ref{rel:work} contextualize our contribution within the current literature, while Section~\ref{sec:data} describes the survey data analyzed in this article; Section~\ref{sec:model} introduces the proposed dynamic Bayesian latent-class regression model, and Section~\ref{sec:application} illustrates the results and empirical findings. Finally, Section~\ref{sec:discussion} provides a brief discussion.

\section{Related work}
 \label{rel:work}

During the past two years, there has been a broad interest in modeling the evolution of the \covid{} pandemic.
Indeed,  incidence data such as the number of positive individuals, hospitalizations, and intensive care unit admissions have been systematically collected and released to the public on regular basis;  for example, in Italy they have been released daily by the Department of ``Protezione Civile'' (\url{https://github.com/pcm-dpc/COVID-19}). 
Accurate modeling of these data has proven useful to measure the current state of the pandemic, to develop strategies based on empirical evidence, and to evaluate the success of different policies adopted by the governments.
 In this context, Girardi et al.\citep{girardi:2020} developed a robust non-linear regression  to model and predict the contagion dynamics of \covid{} in Italy. Alaimo Di Loro et al.~\citep{alaimo:2021} introduced a novel generalized linear model based on Richards' curves, obtaining accurate short-term forecasts of incidence indicators.
Girardi et al.\citep{girardi:2022} also  proposed a change-point growth model that is able to capture subsequent pandemic waves, while Scrucca~\citep{scrucca:2022} developed a real-time index that summarizes the current state of the pandemic. 
 We also refer to\citep{sebastiani:2020, farcomeni:2021, celani:2021, durso2022} for additional modeling strategies and analysis of Italian incidence data. 
 Similar analysis have been developed at the European \citep{cabras:2021, padellini:2022,selinger:2021} and global level. \cite{ihme:2020, reddy:2021,roosa:2020}

Less attention has been devoted to study the impact of \covid{} on individual attitudes, analyzing personal behavior and their interactions with compliance with preventive measures. For example,
Daoust et al.,~\citep{daoust:2020} studied the different level of compliance of Canadian provinces, using composite indicators, and
Krekel et al.~\citep{krekel:2020} studied the associations between happiness and level of compliance with government regulations. Behavioral data have also been used as important predictors of the number of \covid{} cases.~\citep{tripathy:2021}
In Italy,
Duradoni et al.~\citep{duradoni:2021} studied the psychological profile of people that were compliant with the government regulations a month after the lockdown started,  while Guazzini et al.~\citep{guazzini:2022} tested changes in the psychological adaptation across the first two waves of the pandemic.
In this context, we consider the evolution of the compliance of Italian population,  quantifying variations of this compliance  over time, across socio-demographic groups, and studying its associations with the severity of the measures adopted by the Italian government throughout the pandemic.

\section{Data description}
\label{sec:data}

\begin{table}
\caption{\label{tab:dataset} 
	List of analyzed survey  items with code, label, and description.  Subjects  can respond to each item  with ``Not at all'', ``Rarely'', ``Sometimes'', ``Frequently'' and ``Always'', according to their level of agreement with each survey item.}
	\centering
\resizebox{0.98\textwidth}{!}{  
	\fbox{%
\begin{tabular}{*{3}{l}}
 Survey Code & Label & Description \\
  \midrule
 \texttt{i12\_health\_1}     & \textsc{ih1}  & Worn a face mask outside your home (e.g.  when on public transport going\\
		             &               & to a supermarket or going to a main road) \\
 \texttt{i12\_health\_2}     & \textsc{ih2}  & Washed hands with soap and water \\
 \texttt{i12\_health\_3}     & \textsc{ih3}  & Used hand sanitiser \\
 \texttt{i12\_health\_4}     & \textsc{ih4}  & Covered your nose and mouth when sneezing or coughing \\
 \texttt{i12\_health\_5}     & \textsc{ih5}  & Avoided contact with people who have symptoms or you think may have been  \\
		             &               & exposed to the coronavirus \\
 \texttt{i12\_health\_6 }    & \textsc{ih6 } & Avoided going out in general \\
 \texttt{i12\_health\_7 }    & \textsc{ih7}  & Avoided going to hospital or other healthcare settings \\
 \texttt{i12\_health\_8 }    & \textsc{ih8}  & Avoided taking public transport \\
 \texttt{i12\_health\_11}    & \textsc{ih11} & Avoided having guests to your home \\
 \texttt{i12\_health\_12}    & \textsc{ih12} & Avoided small social gatherings (not more than 2 people) \\
 \texttt{i12\_health\_13}    & \textsc{ih13} & Avoided medium-sized social gatherings (between 3 and 10 people) \\
 \texttt{i12\_health\_14}    & \textsc{ih14} & Avoided large-sized social gatherings (more than 10 people) \\
 \texttt{i12\_health\_15}    & \textsc{ih15} & Avoided crowded areas \\
 \texttt{i12\_health\_16}    & \textsc{ih16} & Avoided going to shops \\
\end{tabular}}}
\end{table}

We analyze repeated cross-sectional survey data provided by the Institute of Global Health Innovation (\textsc{ighi}) at the Imperial College of London, in collaboration with the company YouGov. \citep{jones2020imperial} 
Data are publicly available for research purposes at the repository~\url{https://github.com/YouGov-Data/covid-19-tracker}, along with a brief description of the collected variables.
This survey aims to investigate how different populations responded to \textsc{covid-19}, gathering self-reported data on several aspects of the pandemic, including objective measurements, such as testing results and observed symptoms, and subjective measurements, such as daily behaviors.

We are interested in subjective measurements describing the compliance with national preventive regulations. 
This measurements include questions on self-isolation, avoidance of social gatherings, frequency of hand washing, use of hand sanitizer and contact with other people, among many others. 
The complete list of variables used in this study is reported in Table~\ref{tab:dataset}, which includes $14$ out of the $20$ subjective measurements available in the survey. 
The removed items were considered uninformative for the population behavior in relation with the \covid{} measures adopted by the Italian government. 
For example, one removed question asked whether children living in the same household were going to school; however,   schools have been closed for most of the pandemic and students could only attend lectures  remotely.

Publicly available data focus on $T = 37$ survey waves conducted from April 2020 to July 2021.
Notably, the survey waves are not administered at regular time-intervals; for example, $4$ survey waves were submitted in April and May 2020, and only $1$ in July and August 2020; see Table~\ref{tab:dates} for details on the waves administration dates.
Without loss of generality, we will indicate this time grid with $\mathcal{T} = \{t_1, \dots, t_T\}$.
In each wave $t \in \mathcal{T}$, cross-sectional data were collected for a representative sample of $1000$ statistical units, indexed as $i=1,\dots ,n_t = n$.
Each unit is associated with a sampling weight $w_i(t)$ and a vector of covariates $\bx_i(t) = [x_{i1}(t), \dots, x_{im}(t)]^\intercal$, including age, sex, region of residence (``North-West'', ``North-East'', ``Center'', ``South'', ``Islands'') and employment status (``Full-time employment'', ``Part-time employment'', ``Not working'', ``Student'' and ``Retired''). Note that unit $i$ at wave $t$ indicates a different subject than unit $i$ at wave $t^\prime \neq t$; for this reason, we include the wave index $t$ in the covariate vector as $\bx_i(t)$, even if the covariates $\bx_i(t)$ can be considered fixed over time. Refer to Jones~\citep{jones2020imperial} for further information.

This data collection scheme is commonly referred to as ``repeated cross-sectional'' or ``pseudo-panel'' in the literature. We refer to Verbeek~\citep[]{verbeek:2008} for an extensive discussion on the advantages and limitations of  repeated cross-section data compared to  longitudinal data. In longitudinal data, temporal variations  can be modeled, for example, relying on Hidden Markov Models; refer to \textit{Latent Markov models for longitudinal data}~\cite{bartolucci:2012} and references therein for more details. 
These approaches model the transitions across ideal behaviors for each individual, and allow to include the effects of covariates directly in the transition probabilities that characterize the time dynamic.
In our application, we cannot characterize individual trajectories due to the cross-sectional nature of the survey data. 
This is in fact a major limitation of cross-sectional data along with their  inability to monitor changes in personal attitudes. On the other end, repeated cross-sectional data allow to effectively model  average tendencies of a population across time and, since they are less effected by attrition, they typically present a larger sample sizes than longitudinal data. \citep[]{verbeek:2008}
Therefore in Section~\ref{sec:model}, we describe a dynamic regression model for repeated cross-sectional data that can characterize
average variations of the attitude of Italian population during the pandemic.

\begin{table}
	\caption{\label{tab:dates} Calendar date of the 37 analyzed  survey waves.}
\centering
\fbox{%
\resizebox{\textwidth}{!}{%
\begin{tabular}{*{9}{c}}
Wave & 1 & 2 & 3 & 4 & 5 & 6 & 7 &  8  \\
Date & 2 Apr 2020& 8 Apr 2020& 16 Apr 2020& 26 Apr 2020& 1 May 2020& 7 May 2020& 13 May 2020 & 29 May 2020 \\
\midrule
Wave & 9 & 10 & 11 & 12 &13 & 14 & 15 &  16  \\
Date & 10 Jun 2020& 25 Jun 2020 & 8 Jul 2020& 23 Jul 2020& 7 Aug 2020& 19 Aug 2020& 3 Sep 2020& 12 Sep 2020 \\
\midrule
Wave & 17 & 18 & 19 & 20 &21 & 22& 23 &  24  \\
Date&  2 Oct 2020& 15 Oct 2020 & 28 Oct 2020& 11 Nov 2020 & 16 Dec 2020& 6 Jan 2021& 13 Jan 2021& 27 Jan  2021\\
\midrule
Wave & 25 & 26 & 27 & 28 &29 & 30& 31 &  32  \\
Date & 10 Feb 2021& 24 Feb 2021 & 10 Mar 2021& 24 Mar  2021& 7 Apr 2021&  21 Apr 2021 & 5 May 2021 & 19 May 2021 \\
\midrule
Wave & 33 & 34 & 35 & 36 &37 & &  &    \\
Date & 2 Jun  2021& 16 Jun 2021 & 30 Jun 2021& 14 Jul 2021& 28 Jul 2021&   &  &\\
\end{tabular}}}
\end{table}

\section{Methods}\label{sec:model}
\subsection{Model specification and interpretation}
Let $\mathbf{y}_i(t) = (y_{i1}(t), \dots, y_{ip}(t))^\intercal$ denote the responses of subject $i$ to the survey items outlined in {Table~\ref{tab:dataset}} during survey wave $t$; without loss of generality, we encode these responses as $y_{ij}(t) \in \{1, \dots,d\}$ for $j=1,\ldots,p$, with $d=5$ and $p= 14$.
We indicate with ${\mathbf{y}(t) =\{ \mathbf{y}_i(t), \, i=1,\ldots,n \}}$ the observed responses for the $n$ interviewed subjects during the survey wave $t$.
Our main goal is to study the evolution of the population's behaviors over time and during different stages of the \covid{} pandemic. To accomplish this goal, we specify a low-dimensional dynamic model for multivariate categorical data, that characterizes the time-varying probability mass function 
of $\mathbf{y}_i(t)$ in terms of a set of static latent classes and dynamic class-memberships.
Since at each survey wave we observe $n$ statistical units chosen according to a survey design, we adjust the proposed model likelihood to obtain an estimate of the population parameters.

In order to describe the evolution of $\mathbf y(t)$ over time, we assume that the population can be divided into $H$ ideal behaviors (or profiles) that express different attitudes toward \covid{}.	
These ideal profiles can correspond to different behavioral patterns, such as people that rigidly observed all rules and directives to avoid the spread of the disease, or people that kept behaving as usual, ignoring the emergency.
The interpretation and the structure of the ideal behaviors is considered fixed over time, while their proportion can change dynamically. For example, it is likely that, before seeing the effect of \covid{}, several people were not concerned about using public transport or having guests at home. However, as the disease spread and  the effects became more evident, these inclinations  potentially reversed for a subset of individuals.

\begin{figure}[t]
\centering
\begin{tikzpicture}[scale=1.1, transform shape]
\tikzstyle{main}=[circle, minimum size = 13mm, thick, draw =black!80, node distance = 16mm]
\tikzstyle{connect}=[-latex, thick]
\tikzstyle{box}=[rectangle, draw=black!100]
  \node[main, fill = white!100] (theta) {$\nu_{ih}(t)$ };
  \node[main, fill = black!10] (x) [left=of theta] {$\bx_{i}(t)$ };
  \node[main] (z) [right=of theta] {$z_{i}(t)$};
  \node[main, fill = white!100] (beta) [above=1cm of theta] {$\bbeta_h$};
  \node[main, fill = white!100] (eta) [below=1cm of theta] {$\eta_h(t)$};
  \node[main, fill = black!10] (w) [right=of z] {$\mathbf{y}_i(t)$ };
  \node[main] (psi) [right=of w] {$\boldsymbol{\psi}_{h}^{(j)}$};
      \path        (theta) edge [connect] (z)
        (z) edge [connect] (w)
       (psi) edge [connect] (w)
       (eta) edge [connect] (theta)
       (beta) edge [connect] (theta)
        (x) edge [connect] (theta);
\node[rectangle, inner sep=6.5mm, draw=black!100, fit = (theta) (beta) (eta)] {};
    \node[rectangle, inner sep=6.5mm, draw=black!100, fit = (psi)] {};
        \node[rectangle, inner sep=12mm, draw=black!100, fit = (psi)] {};
  \node[] at (0.1,-3.3){{\small{$h=1, \ldots, H$}}};
    \node[] at (9,-1) {{\small{$j=1, \ldots, p$}}};
        \node[] at (9.50,-1.6) {{\small{$h=1, \ldots, H$}}};
\end{tikzpicture}
\caption{Graphical representation of the mechanism to generate data $\mathbf{y}_i(t)$ from model \eqref{eq:populationParafac}.}\label{fig:population}
\end{figure}
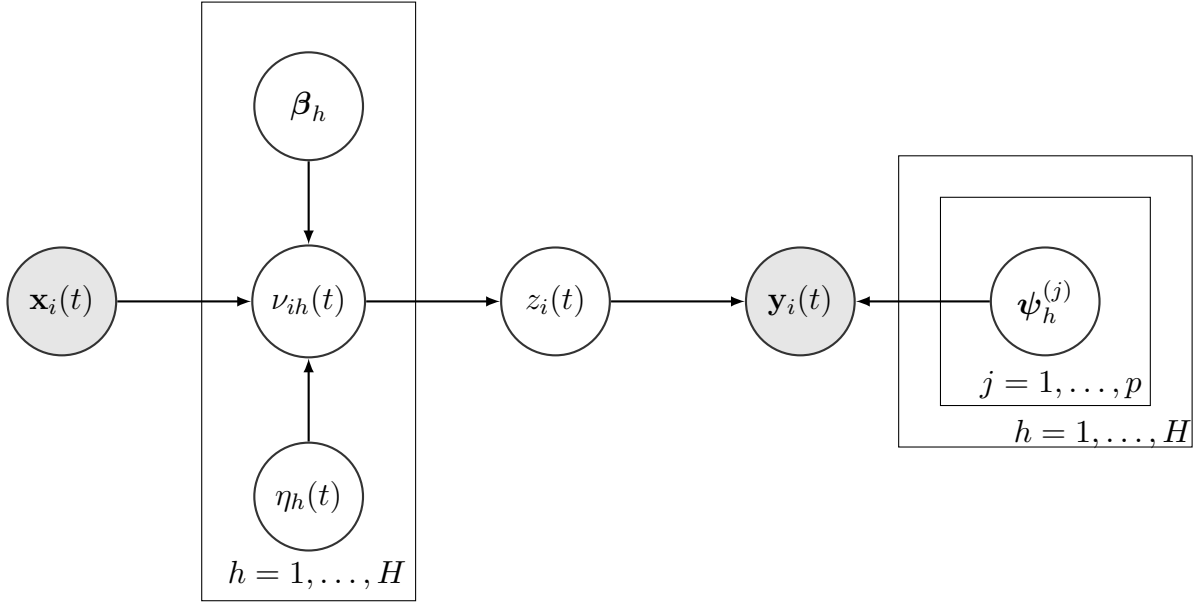

This evolution is modeled relying on a dynamic latent-class regression model, graphically described in Figure~\ref{fig:population}. 
For each wave $t$, subject $i$ is associated to one of the $H$ profiles via the discrete latent variable $z_{i}(t) \in \{1,\dots, H\}$ with probability $\nu_{ih}(t) = \mbox{pr}(z_{i}(t) = h \mid \bx_i(t))$  that depends on the observed covariate vector $\bx_i(t)$. 
Given the profile membership $z_{i}(t) = h$, the probability that subject $i$ responds with $c_j$ to the survey item $j$ is denoted as ${\psi^{(j)}_{hc_j} = \mbox{pr}(y_{ij}(t) = c_j \mid z_{i}(t) = h)}$, with $c_j \in \{1,\dots,d\}$ and $j=1,\dots,p$.
Therefore, we assume that the profiles' attributes are constant over time, and can be characterized by the conditional probabilities of responding with a certain category to the different items, namely
${\boldsymbol{\Psi}_h = \{\psi^{(j)}_{hc_j}, c_j = 1, \dots, d,\, j=1, \dots, p \}}$. 
Within each mixture component, the observed survey items are modeled as a realization from a product of independent multinomial distributions. And, under the considered conditional independence assumption, the dependence structure of the survey items is entirely induced via marginalization of the mixture weights.

Potentially, we could rely on alternative model specifications, exploiting the ordering of  the survey items categories. For example, we could let the difference of probabilities across adjacent categories for each item to be constant, effectively reducing the number of parameters in the model. However, this formulation implies that the distributions of the survey items are stochastically ordered. This assumption might be non-trivial to check empirically, in particular if several items are modeled jointly; therefore we prefer to follow the common practice of using a latent class model for multivariate ordered categorical data, since this model is robust to violations of ordering assumptions.\citep{linzer:2011}

The profile-specific membership varies across the survey waves and according to subject specific covariates. Specifically, the evolution of the profile memberships is modeled trough the probability vector ${\boldsymbol{\nu}_i(t) = (\nu_{i1}(t), \dots, \nu_{iH}(t))^\intercal}$, with $\nu_{ih}(t)$ denoting the probability that subject $i$ of survey wave $t$ belongs to the $h$-th profile and $\sum_{h=1}^H\nu_{ih}(t)=1$.
This parameter is decomposed into two quantities: a profile-specific component $\eta_h(t)$ and a subject-specific effect $\bbeta_h^\intercal \bx_i(t)$ that linearly depends on their observed covariates, with  $\bbeta_h=(\beta_{1h}, \dots, \beta_{mh})^\intercal$.
The dynamic component $\eta_h(t)$ characterizes the temporal evolution of the profiles' memberships, while the linear term for the effect of subject-specific covariates.

Isolating  the individual  and the profile-specific component---which is shared across the population---is particularly relevant in our setting, where we analyze repeated cross-sectional data,  including different subjects across survey waves.
Indeed, this structure allows to estimate the shared dynamic component $\eta_h(t)$, adjusting for the different composition of the cross-sectional population, and measuring the effect of demographic information on the probability of belonging to a certain profile. 

According to these specifications, the model depicted in Figure~\ref{fig:population} can be formalized as
\be
	\mbox{pr}(y_{i1}(t) = c_1, \dots, y_{ip}(t) = c_p \mid \bx_i (t))	&= \sum_{h=1}^H\nu_{ih}(t) \prod_{j=1}^p\psi^{(j)}_{hc_j},\\
	\log\left(\frac{\nu_{ih}(t)}{\nu_{i1}(t)}\right) &= \eta_h(t) + \bbeta_h^\intercal \bx_i(t).
    \label{eq:populationParafac}
    \ee
\noindent
Using the first latent profile as a reference, we let $\eta_1(t) = 0$ and $\beta_{k1} = 0$ for $k=1,\dots,m$ and interpret each $\beta_{kh}$
as the effect of the $k$-th covariate on the log-odds of belonging to profile $h$, instead of the first one, as in multinomial logit regression~\citep[e.g.,][]{azzalini:2012}.
The conditional independence assumption  among the $p$ categorical variables, and the inclusion of the temporal component in the mixture weights, leads to substantial  dimensionality  reduction in the number of model parameters,  while incorporating borrowing of information across survey waves.
The effect of  of time in the model~\eqref{eq:populationParafac} can be made explicit as follows. 
Let $\mathbf{X}$ denote an $(\sum_{t \in \mathcal{T}} n_t \times m)$-dimensional design matrix obtained  stacking the vectors ${[\mathbf{x}_i(t), i=1,\dots, n_t,\, t\in \mathcal{T}]^\intercal}$ by rows;
i.e., the first $n_1$ rows of $\mathbf{X}$ correspond to the observation of the $n_1$ individuals interviewed during first survey wave $t=t_1$, the rows from $n_1 + 1$ to $n_2$ correspond to the individuals interviewed at the second survey wave $t=t_2$, and so on. 
In matrix form, the linear predictor of the second line of Equation~\eqref{eq:populationParafac} can be expressed as
\[
	\mathbf{I}\boldsymbol{\eta}_h +  \mathbf{X}\bbeta_h,
\]
where $\mathbf{I} = \mbox{diag}(\boldsymbol{1}_{n_1}, \dots, \boldsymbol{1}_{n_T})$ is a
$(\sum_{t \in \mathcal{T}} n_t \times T)$-dimensional matrix which identifies the survey wave of each observation, and $\boldsymbol{1}_n$ is a vector of ones with length $n$; similarly, $\boldsymbol{\eta}_h=[\eta_h(t_1), \dots,\eta_h(t_T)]$ is a $T$-dimensional vector containing the values of the dynamic intercepts.
This specification illustrates that time variations in the mean composition of each profile are characterized non-parametrically, since the model includes a different value of the intercept for each survey wave, instead of assuming a parametric relationship between consecutive values of $\eta_h(t)$.

Following Equation~\eqref{eq:populationParafac}, the likelihood contribution for subject $i$ in the survey wave $t$ can be expressed as
\begin{equation}
	\mathbf{p}(\mathbf{y}_i(t) \mid \boldsymbol{\nu}_i(t), \bx_i(t), \boldsymbol{\Psi}) = \sum_{h=1}^H\nu_{ih}(t) \prod_{j=1}^p \prod_{c_j = 1}^{d} \left[\psi^{(j)}_{hc_j}\right]^{\mathds{1}[y_{ij}(t) = c_j]},
  \label{eq:lli}
\end{equation}
where $\mathds{1}[A]$ denotes the indicator function for the event $A$.
To mitigate the effect of potential bias due to the survey design, we follow the approach described by
Vermunt and Magidson~\citep{vermunt:2007}
and Patterson et al.,~\citep{patterson:2002} and incorporate the survey weights into the model via 
\begin{equation}
	L( \boldsymbol{\nu}(t), \boldsymbol{\Psi}) \propto \prod_{i=1}^n \left[\mathbf{p}(\mathbf{y}_i(t) \mid \boldsymbol{\nu}_i(t), \bx_i(t), \boldsymbol{\Psi})\right]^{w_i(t)},
	\label{eq:weightedLikelihood}
\end{equation}
exponentiating each likelihood contribution in~\eqref{eq:lli} to the corresponding survey weight $w_i(t)$. 
The pseudo-likelihood in~\eqref{eq:weightedLikelihood} is used as building-block of several likelihood-based procedures that include survey weights,~\citep{godambe:1986,hesketh:2006, skinner:2012} and recently in some Bayesian methods.~\citep{gunawan:2020} Alternatively, in a Bayesian setting, 
one can use the survey weights to approximate the population generative mechanism, and inferring characteristics of the non-sampled units~\citep{pillai:2015} or, when available, use strong prior information to correct for the sampling bias. \citep{gao:2021}
However, a weighted likelihood approach is conceptually simpler and computationally more efficient when the sampling mechanism is unknown, as in our application.~\citep{gunawan:2020}
     
\subsection{Prior specification and posterior computation}
\label{sec:priorpost}
We consider a Bayesian approach to inference, using the pseudo-likelihood outlined in Equation~\eqref{eq:weightedLikelihood}.  An advantage of this approach is that prior
regularization can avoid convergence issues of maximization algorithms, such as Expectation-Maximization (\textsc{em}),  when used in latent-class regression.\citep{linzer:2011,durante:2019}
Also, a Bayesian approach simplifies modeling of temporal dependencies across survey waves leveraging  a hierarchical dynamic model. 
To effectively estimate  the dynamic intercepts $\boldsymbol{\eta}_h$ inducing borrowing of information across survey waves,
we assume that the variations in profile composition are smooth over time, and leverage a Gaussian Process (\textsc{gp}) prior for the joint distribution of $\boldsymbol{\eta}_h$ over $\mathcal T$; refer, for example,  to
\textit{Gaussian Processes for Machine Learning}\citep{gp:2006} for an introduction to \textsc{gp}.
Recalling that the first group is fixed as a reference, this prior assumes for each latent group $h = 2, \dots, H$
\begin{equation}
	\eta_{h}(\cdot) \sim \mbox{\textsc{gp}}(0,\mathbf{C}_{\vartheta_h}), \quad \mbox{with} \quad  \mathbf{C}_{\vartheta_h}(t,t') = \vartheta_{1h}\exp\left( - \frac{(t-t')^2} {2\vartheta_{2h}} \right) + \vartheta_{3h}\mathds{1}[t=t^\prime],
\label{eq:GP}
\end{equation}
\noindent denoting a squared-exponential covariance function and $\vartheta_{1h} \in \mathbb{R}^+, \vartheta_{2h} \in \mathbb{R}^+$ and $\vartheta_{3h} \in \mathbb{R}^+$ corresponding to the variance, length-scale and noise variance parameters, respectively. \citep{gp:2006}
The squared exponential function favors smooth transitions across time, with the parameters $\boldsymbol{\vartheta}_h = \{\vartheta_{1h}, \vartheta_{2h}, \vartheta_{3h}\}$ controlling the overall structure of the covariance function. Since the covariance is parametrized as a function of the squared time lags across all pairs of time points $(t,t^\prime)$, it accounts for the unequal spacing effectively, inducing higher correlation among closer time points; indeed, continuous stochastic processes such as the \textsc{gp} are appropriate for time series defined over continuous domains, where the spacing across time points is arbitrary, see for example Chapter 6 of \textit{Analysis of financial time series}.\citep{tsay}

Prior specification is completed selecting: independent  log-normal distributions with log-mean $0$ and log-standard deviation $10$ for the components of $\boldsymbol \vartheta_h$, standard Gaussian distribution for the coefficients $\beta_{lh}$, and symmetric Dirichlet distributions for the profile-specific conditional probabilities $\psi^{(j)}_{hc_j}$, letting 
\begin{equs}
   \vartheta_{qh} &\sim \log N(0,10),  \quad q=1, \dots, 3, \quad & h=2,\dots,H, \\
	\beta_{kh} &\sim N(0,1), \quad k=1, \dots, m, \quad & h=2,\dots,H, \\
	(\psi^{(j)}_{hc_1}, \dots, \psi^{(j)}_{hd}) &\sim \mbox{Dirichlet}(1, \dots, 1), \quad j = 1,\dots,p, \quad & h = 1, \dots, H,
\end{equs}
and recalling that $\beta_{hk} = 0$ for identifiability. 

We approximate the posterior distribution of the model parameters via Markov Chain Monte Carlo (\textsc{mcmc}). 
Specifically, we rely on a Hamilton Monte-Carlo algorithm, efficiently implemented in the \textsc{r} package \texttt{rstan},~\citep{rstan} including the weighted likelihood specification of Equation~\eqref{eq:weightedLikelihood}.

\section{Modeling attitude towards Covid-19}\label{sec:application}
To select the number of latent profiles, we evaluated the model performance in predicting the $14$ survey items in Table~\ref{tab:dataset} for various values of $H$, relying on a $4$-folds cross-validation to estimate out-of-sample accuracy.  The data provide evidence that $H=3$ latent profiles provide the best fit  to the data. 
Refer to the Appendix \ref{sec:a1} for further details on model selection.
Posterior inference for the selected model relies on $4000$ iterations collected after a burn-in of $1000$. Convergence was assessed via graphical inspection of the trace-plots, auto-correlation function and convergence diagnostics.
All chains mixed well, with an effective sample size larger than $3500$ for all parameters.
We did not observe label switching across the chains; see Appendix~\ref{app:conv} for further details.
Simulating the $5000$ draws from the posterior required  approximately $4$ hours on a laptop with an Intel i7-7700HQ \textsc{cpu} and $16$GB of \textsc{ram}.

\subsection{Latent profiles description}\label{sec:latentProfile}
\begin{figure}[t]
	\centering
	\includegraphics[width=.95\textwidth]{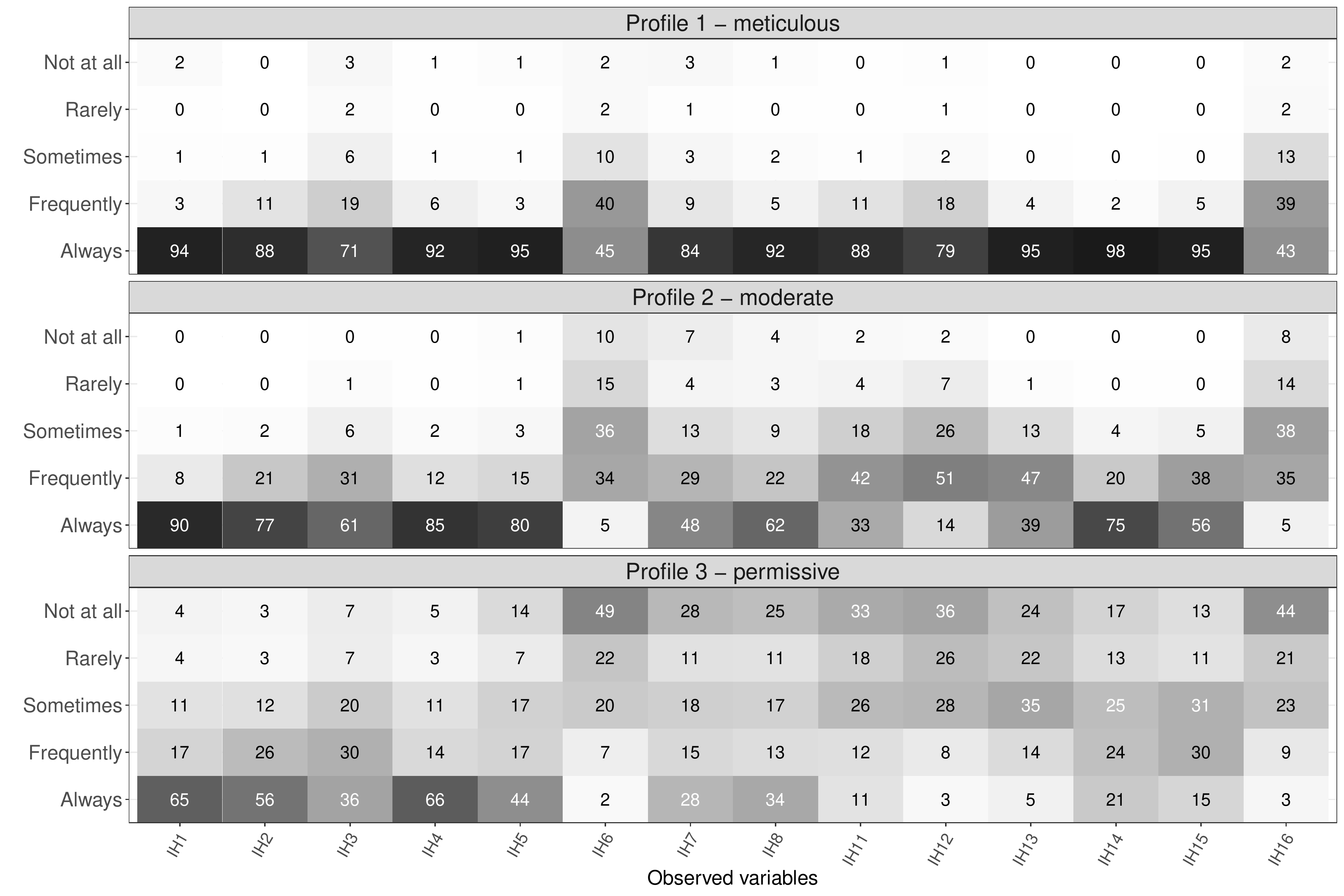}
	\caption{Posterior estimates for the profile-specific parameters $\boldsymbol{\Psi}_h$. The color gradient of the cells varies according to the values of the estimated probabilities, with lighter shades corresponding to smaller values. The number in each cell is the posterior mean of $\boldsymbol{\Psi}_h$ multiplied by $100$. For each item, the  white text corresponds to the response with higher posterior probability.}
	\label{fig:latp}
\end{figure}

In this Section, we describe the composition of the three considered latent profiles (or groups) and their response pattern $\boldsymbol{\Psi}_h$, for $h=1,2,3$, estimated via posterior mean and reported in Figure~\ref{fig:latp}.

The first profile is composed of individuals that scrupulously followed spread-preventive measures. Individuals in this group, with a probability of approximately $90\%$, always wore masks outside home (\textsc{ih1}), washed their hands frequently (\textsc{ih2}), and avoided  taking public transport (\textsc{ih8}).  With probability of $45\%$ they avoided going out in general (\textsc{ih6}). 
Additionally, with a probability of $84\%$ they avoided going to an hospital or health care institution (\textsc{ih7}). Also, they avoided any small gathering (\textsc{ih12}) with probability of $79\%$, and avoided having guests to their home (\textsc{ih11}) with probability of $88\%$ . We will refer to this group as ``meticulous'' in the rest of the article.

The second profile is composed of individuals that moderately followed preventative measure. Compared to the previous group, a smaller fraction of individuals avoided going out (\textsc{ih6}) ($5\%$ probability of responding ``Always''). 
In particular, subjects in this  group completely avoided going to shops (\textsc{ih16}) with probability $5\%$ (compared to the $43\%$ of the meticulous group). A larger fraction of individuals considered small gathering as safe; in fact, the probability of completely avoiding small gatherings (\textsc{ih12}) is of $14\%$, compared with the $79\%$, of the meticulous profile. 
Large gatherings (\textsc{ih14}) are avoided also in this group, with a probability of replying  ``Always'' equal to $75\%$  .
Compared to the meticulous,
the probability of 
taking public transport (\textsc{ih8}) decreased from $92\%$ to $62\%$, and  
the probability of having guests at their home (\textsc{ih11}) decreased from $88\%$ to $33\%$.
Finally, the probability of avoiding to go to an hospital (\textsc{ih7}) decreased from $84\%$ to $48\%$. We will refer to this group as ``moderate''.

The third profile is composed of individuals with a more lenient attitude towards \covid{} measures. In this profile the probability of always wearing a mask outside home (\textsc{ih1}) is $64\%$ compared to $94\%$ and $90\%$ for meticulous and moderate; in particular, the probability of using a mask outside home ``Sometimes'' or ``Rarely'' is $15\%$. 
The probability of responding ``Always'' to  ``Avoided going out in general'' (\textsc{ih6}) is $2\%$ and $49\%$  for the response ``Not at all''.
This profile is characterized by more permissive behaviors also for gatherings. 
For example, the probability of responding ``Not at all'' to items referring to avoiding gatherings is $36\%$, $24\%$ and $17\%$ for small, medium and large gatherings, respectively (items  \textsc{ih12}, \textsc{ih13} and \textsc{ih14}). 
Notably, the probability of responding ``Always'' to ``avoiding crowded places'' (\textsc{ih15}) is only $15\%$, compared to $56\%$ and $95\%$ of the moderate and meticulous profile.  We refer to this group as ``permissive''.

To summarize, the three estimated latent profiles can be interpreted in terms of level of compliance with \covid{} preventive measures.  Some behaviors are similar in the three profiles; for example the modal class for
``Avoided going to the hospital or other healthcare setting'' (\textsc{ih7}) is ``Always'' in all the profiles, even if the distribution of the responses is different across groups. Other behaviors switch across profiles;
for example  in the item ``Avoiding going out in general'' (\textsc{ih6}), we observe
modal category ``Always'', ``Sometimes'' and ``Not at all''  for the meticulous, moderate, and permissive group, respectively.

\subsection{Effects of covariates}
\begin{figure}[t]
	\centering
	\includegraphics[width=.8\textwidth]{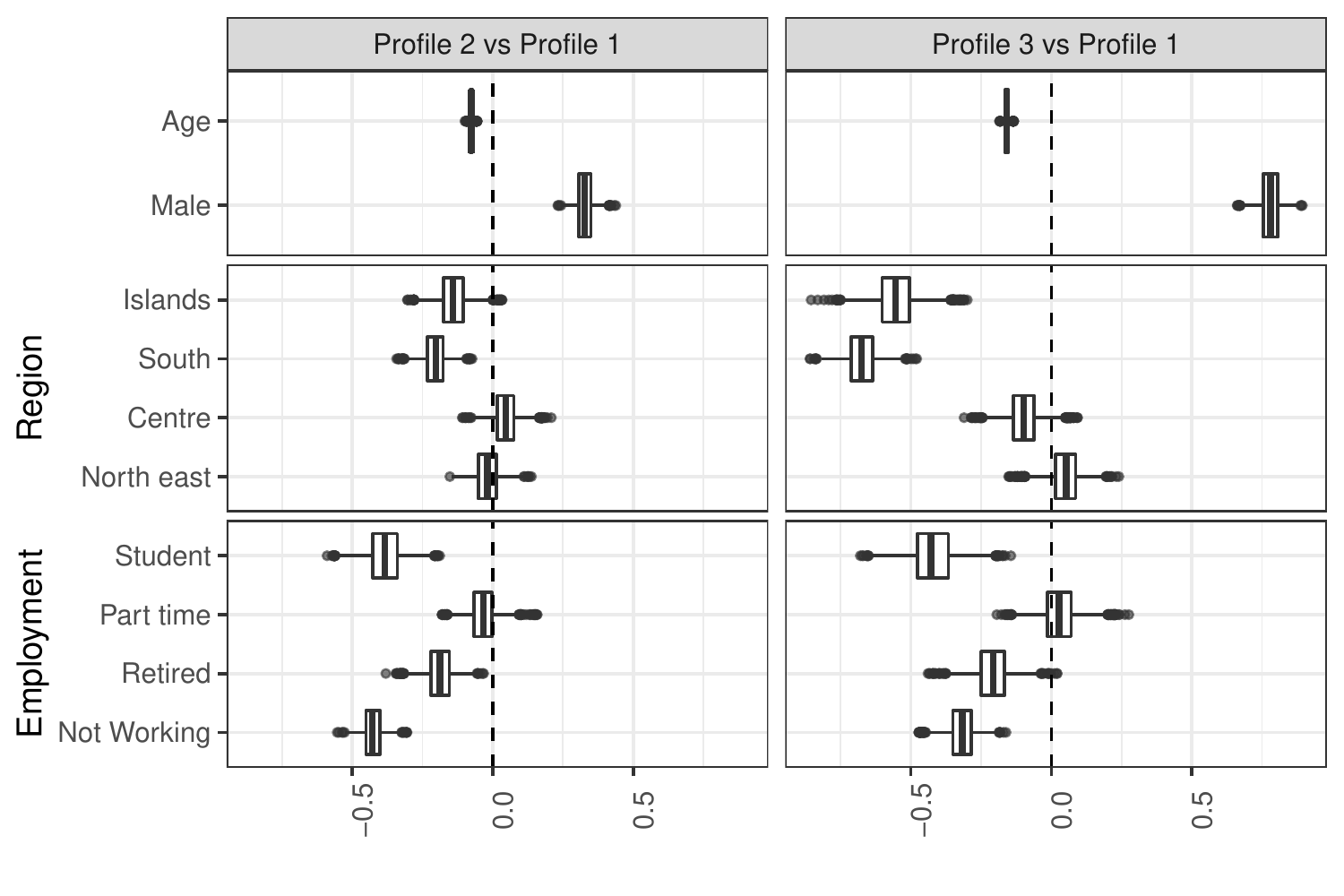}
	\caption{Posterior distributions of the regression coefficients
	$\bbeta_h$ representing the effect of subject characteristics on the profile memberships.}
	\label{fig:covariatesEffect}
\end{figure}

Figure~\ref{fig:covariatesEffect} illustrates the effects of the covariates on the log-odds of  being  assigned to the moderate or permissive group against the meticulous group, that is used as baseline. 
Our empirical findings suggest that older respondents are more likely to be
in the meticulous group rather than in the moderate or permissive ones, since the coefficient for age is negative. For example, the odds of a subject belonging to profile $2$ (moderate) instead of $1$ (meticolous) decrease $\exp(-0.08) = 0.92$ times per each $5$ years of age, while for profile $3$ (permissive) this estimate is $\exp(-0.159) = 0.852$.
Males are more likely to be associated with the permissive groups compared to females; the estimated odds of belonging to profile $2$ instead of $1$ are $\exp(0.328) = 1.388$ times the estimated odds for females.
The odds for profile $3$ instead of $1$ are even larger, with an estimate for males that is $\exp(0.78) = 2.18$ times the estimated odds for females.
We also observe a clear regional effect: individuals living in the south of Italy or in the Italian Islands report higher probabilities to be associated with the meticulous group; for example, the odds of belonging to profile $3$ instead of $1$ for subjects in south Italy and Islands are $\exp(-0.676) =  0.508$  and $\exp(-0.556) = 0.573$ times the estimated odds for North-West, respectively.
Lastly,  students, retired and people that are not currently working show higher probabilities to be associated with the meticulous group rather than moderate or permissive ones.
These results are in line with what reported in 
Carlucci et al.~\citep{carlucci:2020} and references therein, and suggests a more cautious behavior for younger individuals, females, living in south-Italy or Islands and without a full-time occupation.

\subsection{Evolution of the attitude towards \covid{}}\label{sec:profileEvolution}
\begin{figure}[t]
	\centering
	\includegraphics[width=.95\textwidth]{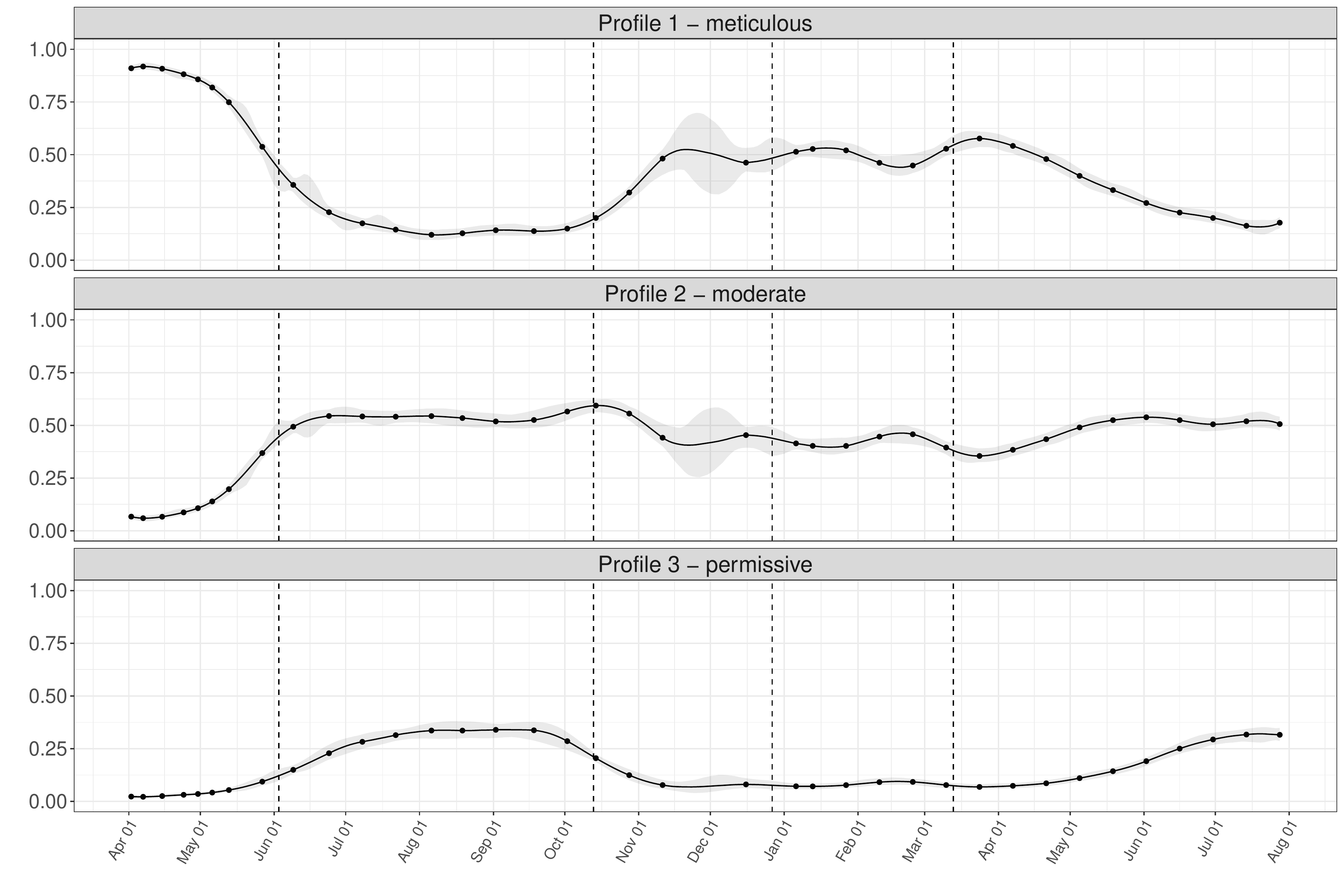}
	\caption{Proportion of Italians associated with the three profiles described in Section~\ref{sec:latentProfile}. Dots indicate the observed waves and grey shaded areas the $95\%$ credible interval.
		Dashed lines correspond to important dates of Italian \covid{} policy and official announcements:
		June 3 2020 the end of the first lockdown;
		October 13 2020 the beginning of the second lockdown;
		December 27 2020 the European vaccine day; and,
		March 13 2021 the day of the publication of the national vaccine plan. 
	}
	\label{fig:dyn}
\end{figure}
To characterize the evolution of the attitudes towards \covid{}, we rely on the properties of the \textsc{gp} 
specification \eqref{eq:GP}, providing  daily predictions
of the proportion of Italians associated with the three profile described in Section~\ref{sec:latentProfile}.
The mean parameters of the \textsc{gp} is approximated predicting the probability to belong to each of the three considered profiles on a new set of locations $\mathcal{T}^\star$  that is a large grid of equally spaced points between the first and last survey wave.  These predictions are mapped to the proportion of Italians associated to each profile described in Section~\ref{sec:latentProfile}, at any given time, by setting the values of the covariates in Equation~\eqref{eq:populationParafac} at their population averages. Posterior estimates are displayed in Figure~\ref{fig:dyn}.

Each panel in Figure~\ref{fig:dyn} is divided in 
five areas, separated by dashed lines that correspond to important dates of Italian \covid{} policy and official announcements. 
The first phase in Italian lockdown lasted from February 21 2020 to June 3 2020, and included several preventative measures such as avoiding leaving the house for non-essential reasons.
In the second phase (June 3 2020--October 13 2020) shops, bars, and restaurants were open to the public, although appropriate social distance was still required.
The third phase (October 13 2020--December 27 2020) correspond to the period of the second Italian lockdown, after \covid{} cases increased over Europe, till the European vaccine day (27 December 2020); this date correspond to the beginning of vaccination policy in Europe.  The fourth period spans from December 27 2020 to March 13 2021, corresponding to the publication of the Italian vaccination plan. Finally, the fifth phase after March 13 2021 correspond to the larger scale diffusion of the vaccine.

According to our analysis, in the interval from April 2 to May 18,  about $86\%$ of the Italian population observed a meticulous behavior, following most of \covid{} preventive measures. In the same period, the proportion of individuals in the moderate profile was $11\%$, and only $3\%$ for the indulgent group. This is in agreement with what reported by
Graffigna et al.
\citep{graffigna:2020}
and Barari et al.~\citep{barari:2020}, although with  different methodologies and data. 

On May 18 2020, the prime minister Conte introduced an easing of the lockdown restrictions, allowing most businesses to open to the public.  Commuting across Regions was still banned until the official announcement of the second phase, on June 3. 
In this period, we notice a rapid variation in the composition of the profiles, with the proportions at June 3 corresponding to $55\%$, $36\%$, and $9\%$ for the meticulous, moderate, and permissive group, respectively.

\begin{figure}[t]
	\centering
	\includegraphics[width=.95\textwidth]{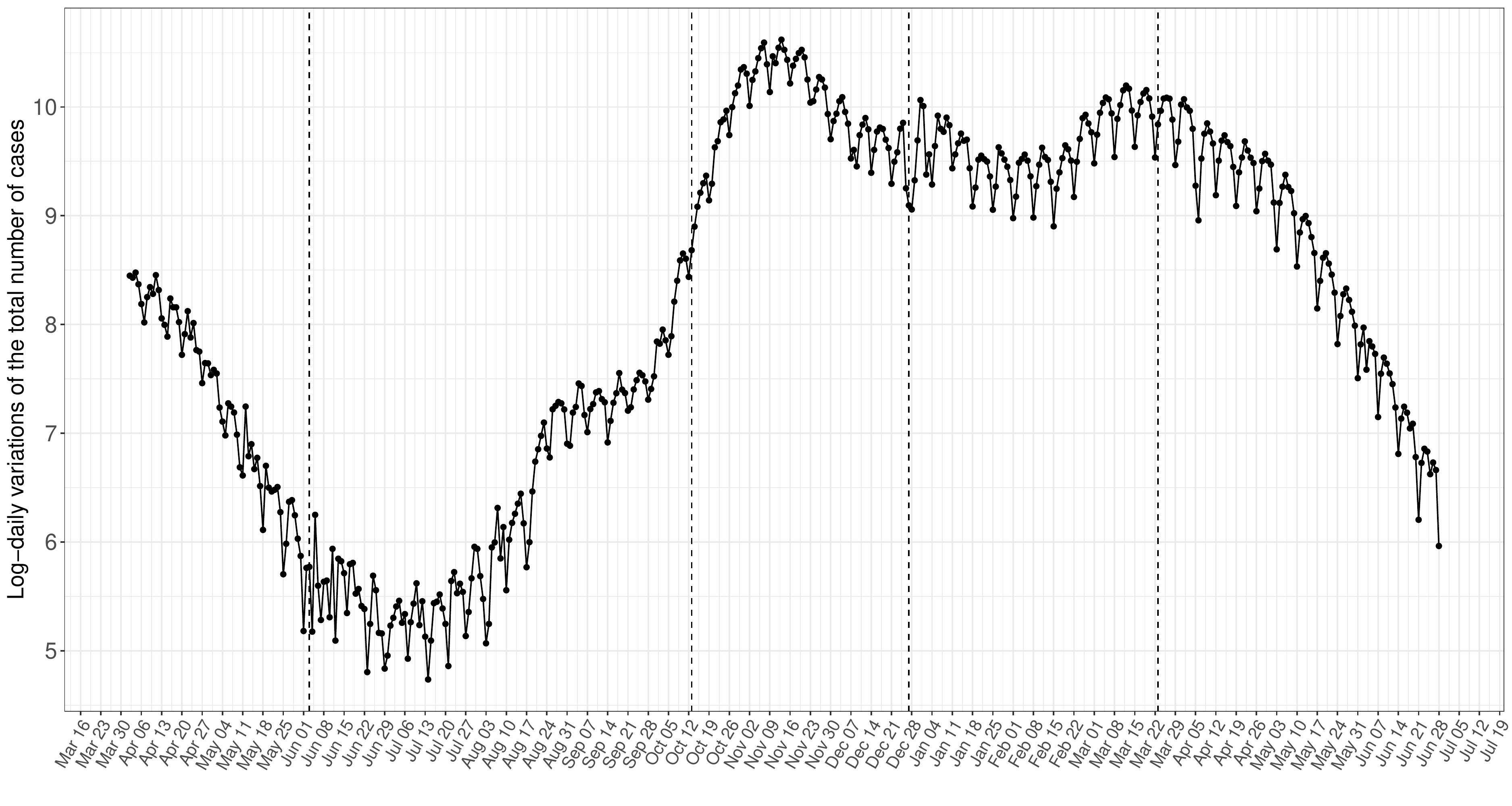}
	\caption{Daily variations in the total number of \covid{} cases in Italy in log-scale.  Dashed lines correspond to important dates of Italian \covid{} policy and official announcements: June 3 2020 the end of the first lockdown; October 13 2020 the beginning of the second lockdown; December 27 2020 the European vaccine day; and, March 13 2021 the day of the publication of the national vaccine plan. The data used for the plot can be downloaded at \url{https://lab24.ilsole24ore.com/coronavirus/}.}
	\label{fig:var}
\end{figure}

In the period ranging from July to October 2020 the proportion of Italians in the three profiles is essentially constant across time, with values close to $17\%$, $54\%$ and $29\%$, respectively. 
In this phase, the number of confirmed cases increased (see Figure~\ref{fig:var}), and half of the Italian population has a moderate attitude towards \covid{}. Also, almost one third of the population is in the permissive group. However, it is worth noting that most of their behaviors, such as not wearing a mask outside, were allowed in this period.

After a new lockdown was enforced on October 13 2020, we observe significant changes in the proportions of Italians associated with each profile.
For example, the proportion of Italians in the meticulous group in the last three observed waves of the third period (October 28, November 11, and December 16 2020) correspond to 
$32\%$, $48\%$, and $46\%$. 
For the moderate group we observe values of 
$56\%$, $44\%$, and $45\%$
while the permissive drop to
$12\%$, $8\%$, and $8\%$.
It is worth noting that despite the reported number of cases of \covid{} on November 11 was comparable to the early phase of the pandemic, only half of the population shows a meticulous  attitude toward \covid{}, compared to $86\%$ of the first lockdown. The permissive group also presents an higher proportion: $8\%$, compared to $3\%$ of the first phase.  These results suggest that, although the increase in the number of cases was comparable between November and April 2020, the behaviors and the reactions of the population were  different, with the second phase characterized by less meticulous behaviors than the previous lockdown.  A possible explanation involves different level of awareness on the disease compared to the first lockdown. For example, in the first period of the pandemic, it was recommended to clean streets and surfaces with disinfectant to avoid the infection; this practice was later flagged as an exaggerate behavior.~\citep{goldman:2020}

The profiles composition stabilizes after the vaccine day on the 27 December 2020, with averages 
for the period December 27 2020--March 13 2021 of $50\%$, $42\%$ and $8\%$ for the meticulous, moderates, and permissive groups, respectively.
This composition remained essentially constant until vaccine became largely availability to the public. In the last phase (after March 13 2021) we can see how the proportion of Italians following a meticulous behavior steeply declined, with a value of $18\%$ for the last wave. Contemporary, the proportion of permissive increased with a value at last wave of $32\%$. The proportion of moderates stabilized around $50\%$ in this phase. This evolution suggests confidence in the vaccine effectiveness.

\section{Discussion}\label{sec:discussion}

We analyzed Italian attitude towards \covid{} leveraging a dynamic Bayesian latent class regression model for survey data. 
The estimated latent profiles can be interpreted as  different degrees of precaution that determine the compliance with the national rules. At the population level, the proportion of Italians associated with each of these profiles follows the various phases of the lockdown. This suggests that Italians have followed the national rules.

There are several potential future directions for this work.
When more recent data will be released, it would be interesting to analyze how the composition of the profiles changes in relation to the  diffusion of new \covid{} variants, such as Omicron.
A further research direction would be to simultaneously study the evolution of attitude towards \covid{} in several countries, comparing different nations and highlighting the main differences.
However, such an extension is not immediate, since the interpretation of the profiles in different country might be substantially different and highly influenced by cultural aspects.

\section*{Appendix}
\appendix

\section{Model selection}
\label{sec:a1}
The number of latent profiles $H$ is selected via $4$-folds cross validations. 
Specifically, for each wave $t\in\mathcal{T}$ we divided data into $4$ folds of equal size, using in turn $3$ folds for estimating the model and the remaining one for evaluating its performance.
Since subjects are different across each wave, folds can be constructed
with random sampling, independently across waves.
We simulate $4000$ samples from the posterior  after a burn-in of $1000$, letting the number of groups $H\in \{2,\dots,8\}$ for each cross-validation fold. Predictive probabilities for each individual and modalities  
 are computed as in Equation~\eqref{eq:lli}, using Monte-Carlo integration, while
 the final out-of-sample predictions have been obtained selecting the modality with largest probability.
We compare observed and predicted values in terms of accuracy, averaging the results in the four cross-validation folds.

Results are reported in Table~\ref{tab:cv}, divided for the $14$ response items. Accuracy increases moving from a model with $H=2$ latent profiles to a model with $H=3$ profiles for most items ($10$ out of $14$), while the remaining items are predicted with the same accuracy by the two models. The prediction accuracy stabilizes with the model with $H=3$ and does not increase further with $H$. Therefore, model with $H=3$ is the most parsimonious model with best fit to the data, and should be selected. 

\begin{table}[t]
	\centering
	\caption{Accuracy, multiplied by $100$. Largest values are highlighted in bold-face, and correspond to the best model. In case of multiple maxima, the smallest model is preferred and the corresponding accuracy is highlighted in bold-face.}
	\label{tab:cv}
	\fbox{%
		\begin{tabular}{crrrrrrr}
			Item label & $H=2$ & $H=3$ &  $H=4$ &  $H=5$ &  $H=6$ &   $H=7$ &   $H=8$ \\
			\toprule
			\textsc{ih1}    & \textbf{78} & 78 & 78 & 78 & 78 & 78 & 78 \\
			\textsc{ih2}     & 61 & \textbf{62} & 62 & 62 & 62 & 62 & 62 \\
			\textsc{ih3}     & 57 & \textbf{58} & 58 & 58 & 58 & 58 & 58 \\
			\textsc{ih4}     & 67 & \textbf{67} & 67 & 67 & 67 & 67 & 67 \\
			\textsc{ih5}     & 65 & \textbf{66} & 66 & 66 & 66 & 66 & 66 \\
			\textsc{ih6 }  & 56 & \textbf{57} & 57 & 57 & 57 & 57 & 57 \\
			\textsc{ih7}   & 58 & \textbf{59} & 59 & 59 & 59 & 59 & 59 \\
			\textsc{ih8}   & \textbf{78} & 78 & 78 & 78 & 78 & 78 & 78 \\
			\textsc{ih11}  & 61 & \textbf{62} & 62 & 62 & 62 & 62 & 62 \\
			\textsc{ih12}  & 57 & \textbf{58} & 58 & 58 & 58 & 58 & 58 \\
			\textsc{ih13}  & \textbf{67} & 67 & 67 & 67 & 67 & 67 & 67 \\
			\textsc{ih14}  & 65 & \textbf{66} & 66 & 66 & 66 & 66 & 66 \\
			\textsc{ih15}  & 56 & \textbf{57} & 57 & 57 & 57 & 57 & 57 \\
			\textsc{ih16}  & \textbf{59} & 59 & 59 & 59 & 59 & 59 & 59 \\
	\end{tabular}}
\end{table}

\section{Convergence diagnostic} \label{app:conv}
\begin{figure}[h!]
	\includegraphics*[width=\textwidth]{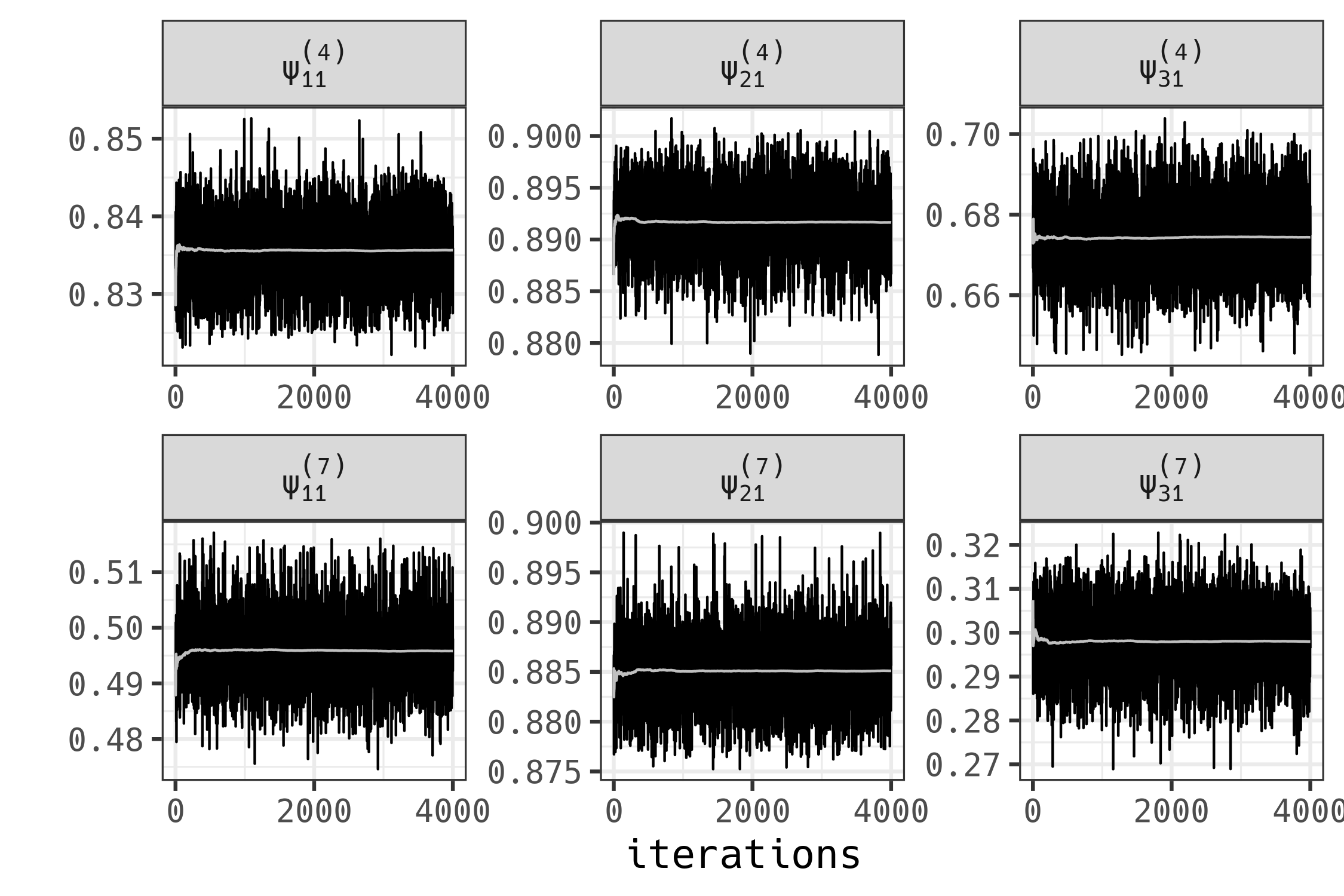}
	\includegraphics*[width=\textwidth]{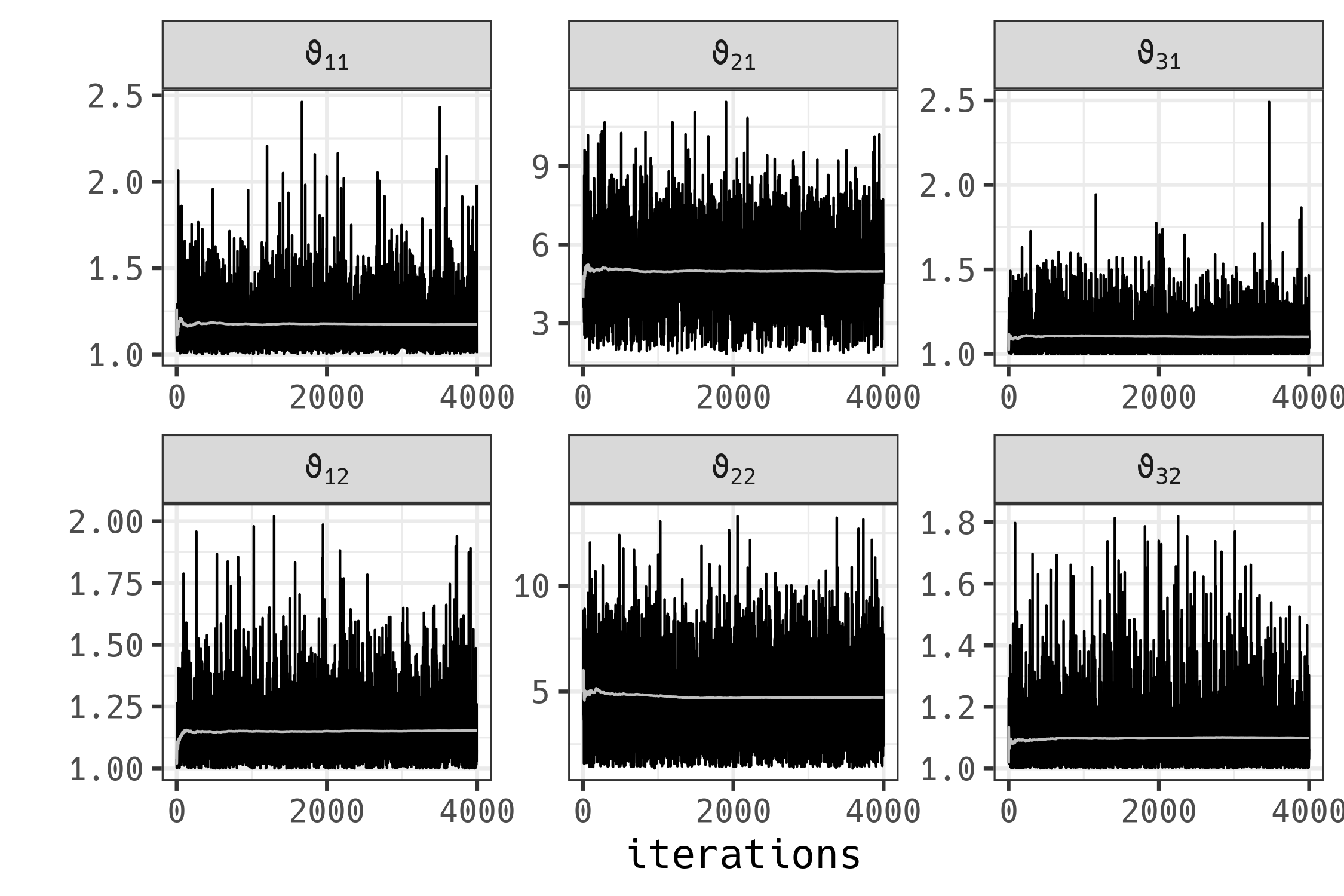}
	\caption{Traceplots of the posterior distribution of some illustrative parameters. Light gray lines denote the 
	cumulative means of the parameters at each iteration.}
	\label{fig:traceplots}
\end{figure}

Figure~\ref{fig:traceplots} shows traceplots of the marginal posterior distributions for some parameters  of the model described in Section~\ref{sec:model}.
The  chains exhibit good mixing,   showing  no  jumps  between  profiles,  that  would  indicate  label  switching; traceplots  for  the remaining model parameters  behave  similarly.

\clearpage
\bibliographystyle{WileyNJD-AMA}
\bibliography{references}

\end{document}